# ANDROID NOTE MANAGER APPLICATION FOR PEOPLE WITH VISUAL IMPAIRMENT

Gayatri Venugopal

Symbiosis Institute of Computer Studies and Research (SICSR),
Symbiosis International University (SIU), Atur Centre,
Gokhale Cross Road, Model Colony, Pune – 411 016, Maharashtra State, India

*ABSTRACT*

*With the outburst of smart-phones today, the market is exploding with various mobile applications. This paper proposes an application using which visually impaired people can type a note in Grade 1 Braille and save it in the external memory of their smart-phone. The application also shows intelligence by activating reminders and/or calling certain contacts based on the content in the notes.*

*KEYWORDS*

*visually impaired, braille, text-to-speech, android, note manager, smart-phone application, reminder, telephony.*

## 1. INTRODUCTION

Mobile phones have become an inevitable part of our daily lives. It is difficult to think of a day without having our mobile phone by our side. The components that were used for our routine jobs have become obsolete now. For instance, we do not see children playing with pocket video games because the games are now available on smart phones to which they have easy access. Similarly, our to-do list has changed its form from a piece of paper to an application on our smart phone. This paper talks about creating a note manager application. This application is used to create notes that the user who owns the smart phone would like to store in his/her device. A note could be his/her to-do list, important points covered during a meeting, address of a shop etc. Basically it could be any sort of information the user would like to keep with him/her. There are many similar applications available on various mobile application markets [1,2] but a person with no sight may not be able to use applications which were created without keeping in mind the users who are visually impaired. In such applications, the user has to type his/her note using a QWERTY keypad, that is, a set of buttons or keys embedded on the device used for input from the user or an on-screen keyboard, also called a soft keyboard. This would be difficult to use for a person who is not able to see the keys, especially if it is a soft keyboard. Hence, the aim is to design and develop such an application for people with any sort of visual impairment.

While creating notes using this application, the smart-phone is intended to be used as a Braille writing pad. Braille is a method invented by Louis Braille at the age of 15 as mentioned [3], using which a visually impaired person can read and write using the sense of touch. A Braille document is essentially a set of cells, each cell represented in the form of a 2 x 3 matrix of raised and/or flat dots as described by Gikes, Cowens and Taylor [4]. A Braille sheet may contain patterns of dots based on one of the three different grades in Braille – Grade 1, Grade 2 or Grade 3 wherein Grade 1 Braille contains a pattern for each character; for instance "a" will correspond to one pattern, "b" will correspond to a different pattern and so on, Grade 2 Braille and Grade 3 Braille use contractions, that is, each word is represented by one pattern as depicted [5]. The application specified in this paper makes use of Grade 1 Braille. The user will be able to create and read notes using the touch and vibration features of the device thus eliminating the need of viewing the contents on the screen. The user can create a note, save it in the device and later, he/she can choose a note and read it. The user may choose from two options to read the note. She/he may listen to the contents of the note by using the speech feature of the device, or by using the touch and vibration feature of the device.

Android platform was chosen to develop the application. Android is an open source platform [6] for mobile devices, which means the code for the platform is available for all, it can be modified, re-used and distributed as explained by Perens [7]. It is a software stack that comes with a Linux kernel, libraries, application framework and applications as stated by Rogers, Lombardo, Mednieks and Meike [6]. A group of mobile technology companies called the Open Handset Alliance (OHA), work to promote Android and its openness [8]. The first commercial version of Android was released in the year 2008. While writing this paper, only five years later, there has been an enormous growth in the use of Android devices and subsequently, Android applications. As of 2013, according to a recent report from IDC [9], Android devices occupy 79.3% of the smart-phone market. As of 2013, Google Play, the official repository for Android applications, contains one million applications as reported [10]. An Android application can be developed on any platform, Windows or Linux, and the developer need not pay any fee to code, test and deploy an Android application on a device. Owing to these reasons, Android platform has been selected to develop the application.

To summarize, the application intends to popularize smart-phone usage among people who may face problems using the device due to inaccessibility of applications.

## 2. DESIGN

The Android software stack contains several layers, the topmost layer being the Applications layer [6]. This is the layer where the application will be installed. The application has been designed for touch-screen devices, keeping in mind the tactile nature of Braille. The screen of the device is divided in six parts, representing one Braille cell containing six dots arranged in the form of a 2 x 3 matrix as shown in figure 1.

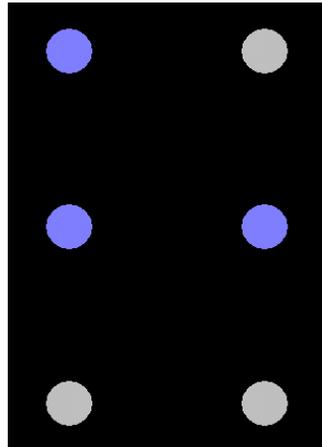

Figure 1. Dots depicting a Braille cell representing the character 'h' in Grade 1 Braille

The vibration feature of the device is used by the application which will cause the device to vibrate when the user touches a dot or any text on the screen. When the device vibrates, it also reads out the component that the user touched thus implementing the text-to-speech (TTS) functionality available on the Android device. The user may choose to turn the voice assistance off once he/she starts using the application regularly. The options in the entire application have been placed in such a way that the user is able to identify them with ease, with the help of vibration and the text-to-speech feature as shown in figure 2. A brief design of the application is depicted in figure 3. The application consists of three options on the main screen, namely "Notes", "Settings" and "Help". "Notes" is used to either compose a new note or to open an existing note. "Settings" is used to turn on or off the text-to-speech feature for the application. It is also used to select whether the user would like to enter the file name in which the content will be saved, in Braille, that is, by activating the dots, or by manually typing it. A visually impaired person who may be able to see partially may use the latter option. The "Help" option is used to train the user in identifying the pattern for each character, with the help of vibration and voice assistant.

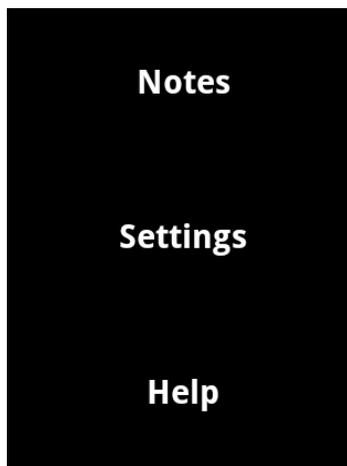

Figure 2. Arrangement of options in the application

A brief design of the application is depicted in figure 3. Android provides APIs (Application Programming Interface) [11] using which we can use the Telephony Manager and the Calendar. An API is a set of classes and methods, with the help of which an application can access the features and functionality of another application or platform. In this case, our application can access the device's telephony feature and the calendar. The Telephony Manager is used to make a call, and the Calendar is used to store and display a reminder to the user. The implementation of these components will be described in the subsequent section.

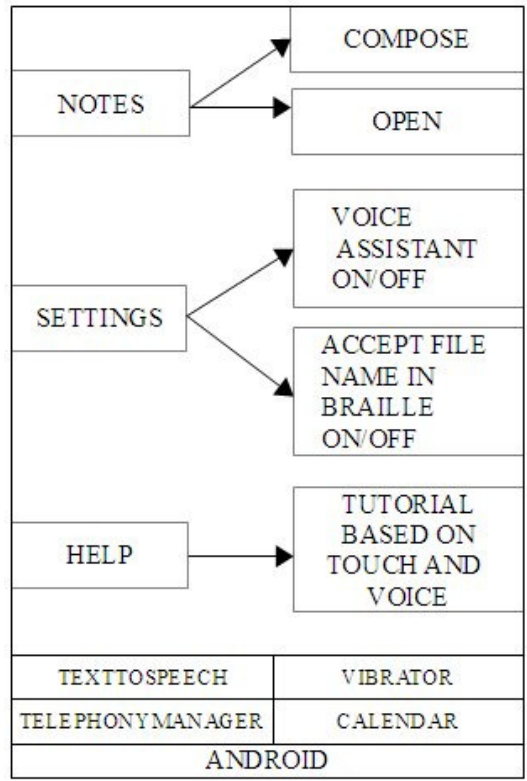

Figure 3. An overview of the application

## 3. WORKING

When the application is launched, it checks if a speech synthesis engine is present in the device. This engine is required to render text and convert it to speech [12]. The content written by the user is expected to be read aloud by the speech synthesis engine in the device. If it is not present, the application redirects to the corresponding download page on Google's Play Store [13], from where the user can download and install the library. The first screen of the application provides the user with three options, which are "Notes", "Settings" and "Help". The screen in this scenario is divided into three horizontal sections. When the user touches the screen, based on the region where the user touched, the voice assistant would read aloud the description for that region. To select an option, the user can double tap anywhere in that region. This directs the user to the appropriate screen. To go back to the previous screen, the user can use a "fling" action, that is, swiping the finger across the screen. To go back, the user can swipe from right to left and to go to the next screen, the swiping action can be used in the opposite direction. Whenever the user touches a region that has a meaning associated with it, the device would vibrate and the voice assistant would describe it, provided this feature has not been turned off by the user from the "Settings" option. This makes it easier to locate dots on the screen. To select a dot, the user is required to press it for some time, commonly referred to as a "long press" event. The voice assistant will guide the user throughout the process, describing and reporting the option or the dot that was touched, whether the operation was successful or not etc.

The cells are numbered from 1 to 6 as is the convention [14], in a top-to-bottom manner as shown in figure 4.

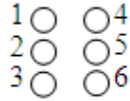

Figure 4. Braille Cell

Once the user selects a pattern of dots in a cell, he/she can swipe from right to left to type the next character in Braille. When the user moves from one pattern to another, the previous pattern is mapped to a Braille-to-English chart as depicted [15], and the pattern that the user selected, is recognized as an English character provided the pattern is valid. Once he/she has finished typing the note, he/she is required to tap on the screen thrice to save the contents in a file. Depending on the mode that the user selected in from the "Settings" option, the user would be prompted to enter the file name. If Braille mode is on, the user can enter the file name in the same manner that he/she entered the content. If it is off, the user can enter text directly. To open a previously saved note, the user can double tap on "Open", which would display a list of files saved by the user. The voice assistant would read aloud the name of the file over which the user keeps his/her finger. As specified earlier, the device would vibrate when the user touches a file name in the list. To open the file, he/she can double tap on that region and the application would then ask the user if she/he would like to read the contents of the file with the sense of touch or using the sense of hearing. In the former case, the contents would be displayed on the screen with selected (or coloured) and plain dots. When the user touches a selected dot, the device will vibrate for a longer duration as compared to the duration of vibration when he/she touches a plain dot. In the latter case, the voice assistant would read aloud the contents of the file, while at the same time, the content would be displayed on the screen. The user can modify the content if required. The "Settings" option has been described above. The "Help" option is used to guide the user on how to use the application. This training uses the voice assistant along with the vibration feature of the device. It lets the user navigate through the Braille patterns for all the characters one at a time. As mentioned above, when the user touches a plain dot, the device would vibrate for a short duration, and when the user touches a selected dot, it would vibrate for a longer duration. The user can triple tap to go back to the main screen.

The application also showcases its intelligence with the help of a reminder and a calling facility. Basically the application would render the note that the user saved, and check for the words "remind me" and "call" at the beginning of the note, and the word "at", if one of these two words are present at the beginning. If "remind me" is detected, the application would generate a reminder and display and read out the message to the user at the time specified after "at", with the help of Calendar present in all Android devices. Similarly the application would call the specified contact or number at the time specified after "at", making use of Android's Telephony Manager, provided a SIM card is present in the device.

## 4.    CONCLUSION

The note manager application aims to improve the quality of life for people with visual impairments, be it complete blindness or partial blindness. It eliminates the need to remember all the important events or tasks for the day. It also reduces the effort required by persons with visual disabilities to use a normal smart-phone. On-screen keyboards available on a regular smart-phone would require more effort as the keys that the user presses do not have a tactile feedback associated with them. Since the application gives feedback using vibration, and also voice, which is an optional feature, it improves the typing experience for the user. Any user who has a good understanding of basic Grade 1 Braille can use the application. Bright colours have not been used keeping in mind people with colour vision deficiency. Using similar colours for different operations may confuse the user if she/he is not able to differentiate between the colours. Thus the application has been designed to avoid any such confusion. Text-to-speech component in an application has been proven useful to make an application accessible. Users may also choose to turn off this facility through the application, if they do not wish the content in the note to be read aloud, thus keeping in mind the privacy of the user. At a time where the market is booming with smart-phones and Android being one among the frontrunners in the race for the most widely used mobile operating system, the application is expected to reach as many people as possible with the intention of making them more comfortable with smart-phones.

**Author**

**Gayatri Venugopal** is a Research Associate at Symbiosis Institute of Computer Studies and Research (SICSR), a constituent of Symbiosis International University (SIU). She is currently teaching web technologies, and mobile application development on Android. Her areas of interest include mobile computing and assistive technologies in IT.
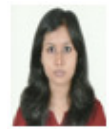